\documentclass[superscriptaddress,aps,prl,twocolumn]{revtex4-1}
\usepackage{graphicx} \usepackage{textcomp} \usepackage{amssymb}
\usepackage{amsmath} \usepackage[utf8]{inputenc}
 \newcommand{\dd}[0]{\mathrm{d}}

\usepackage{amsfonts}

\usepackage{mathbbol}

\usepackage{times} \usepackage{time} \usepackage{mathptmx}

\DeclareSymbolFontAlphabet{\amsmathbb}{AMSb}%

\begin{document}
\title{Active beating of a reconstituted synthetic minimal axoneme}
\author{Isabella Guido} \email{isabella.guido@ds.mpg.de}
\affiliation{Max Planck Institute for Dynamics and Self-Organization
  (MPIDS), 37077 G\"ottingen, Germany} \author{Andrej
  Vilfan}\affiliation{Max Planck Institute for Dynamics and
  Self-Organization (MPIDS), 37077 G\"ottingen, Germany}
\affiliation{Jo\v{z}ef Stefan Institute, 1000 Ljubljana, Slovenia}
\author{Kenta Ishibashi} \affiliation{Graduate School of Frontier
  Biosciences, Osaka University, Osaka 5650871, Japan}
\affiliation{Center for Information and Neural Networks (CiNet),
  National Institute of Information and Communications Technology, Osaka
  565-0871, Japan} \author{Hitoshi Sakakibara} \affiliation{Advanced ICT
  Research Institute, National Institute of Information and
  Communications Technology, Kobe 651-2492, Japan} \author{Misaki
  Shiraga} \affiliation{Graduate School of Life Science, University of
  Hyogo, Hyogo 678-1297, Japan}
 
\author{Eberhard Bodenschatz} \affiliation{Max Planck Institute for
  Dynamics and Self-Organization (MPIDS), 37077 G\"ottingen, Germany}
\affiliation{Institute for Dynamics of Complex Systems,
  Georg-August-University G{\"o}ttingen, 37073 G{\"o}ttingen, Germany}
\affiliation{ Laboratory of Atomic and Solid-State Physics, Cornell
  University, Ithaca, NY 14853, United States} \author{Ramin
  Golestanian} \email{ramin.golestanian@ds.mpg.de}\affiliation{Max
  Planck Institute for Dynamics and Self-Organization (MPIDS), 37077
  G\"ottingen, Germany} \affiliation{Rudolf Peierls Centre for
  Theoretical Physics, University of Oxford, Oxford OX1 3PU, United
  Kingdom} \author{Kazuhiro Oiwa}\affiliation{Advanced ICT Research
  Institute, National Institute of Information and Communications
  Technology, Kobe 651-2492, Japan} \affiliation{Graduate School of Life
  Science, University of Hyogo, Hyogo 678-1297, Japan}

\keywords{Reconstituted axoneme, synthoneme, microtubule pair, axonemal
  dynein, molecular nanomachine}
\begin{abstract}Propelling microorganisms through fluids and moving
  fluids along cellular surfaces are essential biological functions
  accomplished by long, thin structures called motile cilia and
  flagella, whose regular, oscillatory beating breaks the time-reversal
  symmetry required for transport. Although top-down experimental
  approaches and theoretical models have allowed us to broadly
  characterize such organelles and propose mechanisms underlying their
  complex dynamics, constructing minimal systems capable of mimicking
  ciliary beating and identifying the role of each component remains a
  challenge. Here we report the bottom-up assembly of a minimal
  synthetic axoneme, which we call a synthoneme, using biological
  building blocks from natural organisms, namely pairs of microtubules
  and cooperatively associated axonemal dynein motors. We show that upon
  provision of energy by ATP, microtubules undergo rhythmic bending by
  cyclic association-dissociation of dyneins.  Our simple and unique
  beating minimal synthoneme represents a self-organized nanoscale
  biomolecular machine that can also help understand the mechanisms
  underlying ciliary beating.
\end{abstract}
\maketitle
\section{Introduction}

Eukaryotic cilia and flagella accomplish vital physiological
tasks \cite{Satir_Christensen,Gilpin}, ranging from the swimming and
feeding of microorganisms to the clearance of mucous in
airways \cite{Loiseau.Viallat2020}, establishment of left-right body
asymmetry and transport of cerebrospinal fluid in brain
ventricles \cite{Faubel.Eichele2016} and the spinal cord.  Defects in
ciliary assembly or in their function are responsible for diseases that
affect millions of patients worldwide \cite{Fliegauf}. Each cilium's
structural framework is the axoneme, which consists of nine microtubule
doublets and several hundred other proteins that maintain its structure,
drive and control the beat, and transport the materials \cite{Ishikawa}.
The beating is powered by axonemal dyneins that form two arrays (outer
and inner dynein arms) along the side wall of the doublet
microtubules. They generate interdoublet shear that causes the bending
of microtubule doublets \cite{Lin}.

The analysis of \textit{Chlamydomonas} flagellar
mutants \cite{Kamiya2014}, biochemical and biophysical \cite{Oiwa99}
identification and structural studies \cite{Oiwa2003}, especially using
cryoelectron tomography \cite{Lin}, have progressed our understanding on
the molecular organization and function of each axonemal
component \cite{Mitchison2010}.  Recently, axonemal dyneins have been
reported to have mechanical and structural properties so distinct from
other protein motors to be suitable for the fabrication of biomolecular
nanomachine \cite{Oiwa2012, Furuta}.

How this complex structure drives the beating patterns is not yet well
understood \cite{Mitchison2010,Sartori,Lindemann,Lindemann.Lesich2010,Geyer.Howard2016,Sartori.Julicher2016}.
A controlled assembly of minimal systems that reproduce certain aspects
of ciliary dynamics with far less cellular feedback mechanisms can be an
important step towards solving the puzzle of ciliary beating dynamics.
To address this issue we use the bottom-up approach and reconstitute
in-vitro a minimal axoneme made of the main axonemal constituents, outer
dynein arms (ODA) and microtubules, and we reproduce the bending
oscillation of the natural cilia (flagella). We call this structure a
``synthoneme'' in analogy to the the well described axoneme.

The modularity of the reconstituted structure of this synthoneme, its
defined geometry and the nature of the building blocks allow the system
to go beyond the minimal beating systems reported in the past. Namely,
two microtubule doublets extended from frayed and disintegrated axonemes
that exhibit repetitive buckling \cite{Kamiya1986,kamiya2005}; mixtures
of microtubule bundles and kinesin-1 clusters, which self-assemble in
cilia-like beating structures \cite{Sanchez_2011, Vilfan2019}.  The
controlled assembly of our synthoneme has the potential for the
development of controllable beating structures that can serve as a tool
for the understanding of the flagellar/ciliary beating and for eventual
technological advance in the field of bio-inspired systems.
 
\section*{Results}
The reconstituted minimal bending system is made of taxol-stabilized
microtubules and ODA extracted from flagella of the green algae
\textit{Chlamydomonas reinhardtii} (See Supplementary Fig.~S1). We chose
to use ODA for its ability to self-organize in arrays on
microtubules \cite{Owa}, the high yield in preparation, and its robust
motility \textit{in vitro} \cite{Kamiya2014,Furuta2009,Seetharam}.

Microtubules are polymerized in an experimental chamber by flowing
porcine brain tubulin over fragmented \textit{Chlamydomonas} axonemes
(hereafter named seeds). The seeds represent the joint basal end of the
microtubule pair and are spontaneously attached to the coverslip.  This
heterologous seeding procedure provides microtubule pairs with the same
polarity in which the filaments grow at a distance on axoneme
scale. With the seeds the relative movement of microtubules is blocked
at the basal end of the pair, but free at the distal end.

These properties allow the repetitive buckling of the microtubules
driven by dynein. The distance between two polymerized microtubule
singlets at the basal end probably ranges from the minimal distance
between two A-tubules of adjacent doublets (approx.\ 30 nm) and the
maximum corresponding to the diameter of the axoneme (approx.\ 200 nm)
(see Supplementary Fig.~S2).  Since the length of microtubules is longer
than 10 $\mu$m, the tips of the filaments thermally fluctuated unless
ODAs formed cross-bridges between them. Therefore, a precise
determination of the inter-microtubule distance in individual cases is
not applicable in our observations.

\begin{figure*}
  \centering \includegraphics[width=1\textwidth]{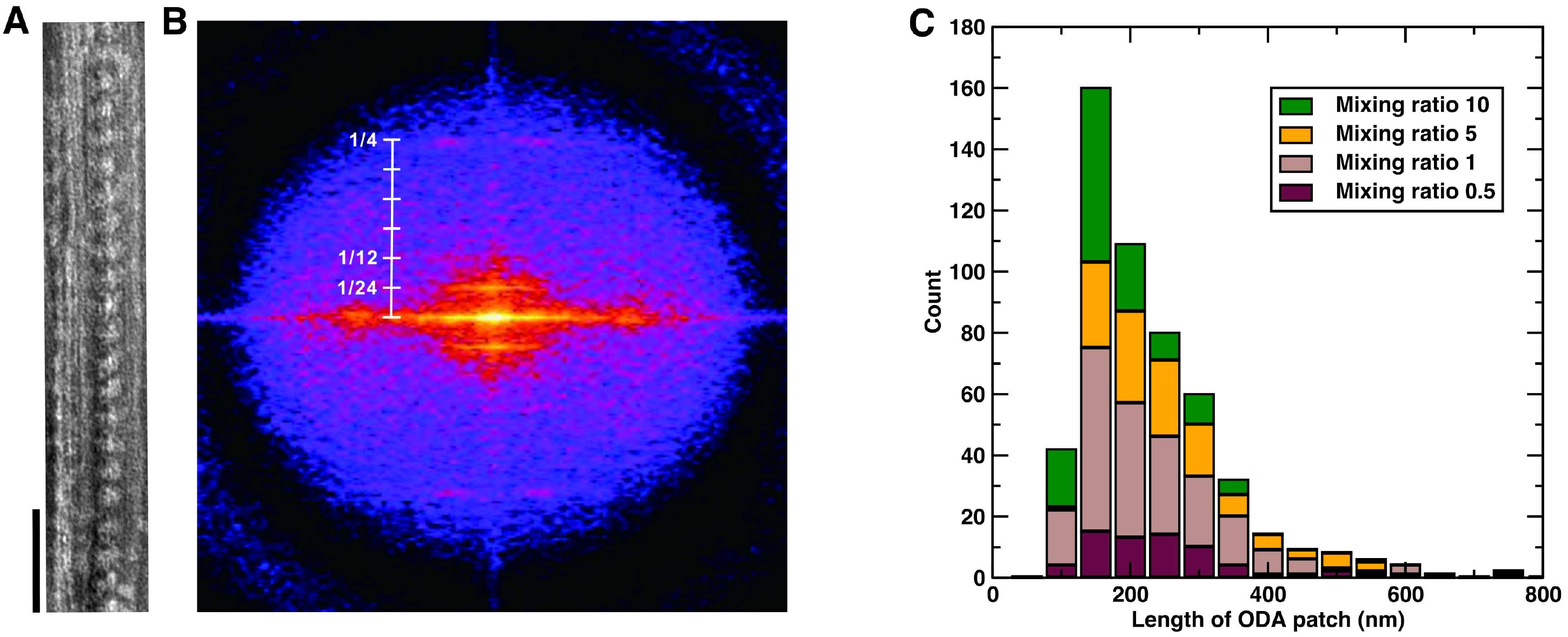}
  \caption{A) Negative staining EM image showing a portion of a
    microtubule microtubule and an assembly of the ODA complex. Globular
    blobs are regularly aligned along the microtubule. Scale bar: 100
    nm. B) Fourier transform of EM images shows a 24 nm periodicity of
    ODA on microtubules.  1/4 nm$^{-1}$ layer lines derived from
    microtubules are clearly observed (scale unit: 1/24 nm$^{-1}$). C)
    Length distribution of ODA patches at different mixing ratios
    between ODA and microtubules. The peak length is around 100--200 nm
    irrespective of the mixing ratio.}
  \label{fig:Fig1}
\end{figure*}

The ODA is provided to the system by perfusion of extract from axonemes
containing ODAs and ODA-docking complex (ODA-DC) into the flow chamber.
Due to the presence of the docking complex, ODAs are associated with the
microtubule with high cooperativity in an end-to-end fashion and thereby
align unidirectionally along one protofilament \cite{Owa}.  We confirm
the reconstitution of ODAs on microtubules with negative staining
electron microscopy (Fig.~\ref{fig:Fig1}A and Supplementary Fig.~S3) and
the longitudinal 24-nm periodicity of ODA/ODA-DC aligned along the
filament is identified by Fourier analysis (Fig.~\ref{fig:Fig1}B). This
periodicity is characteristic for ODAs in native axonemes \cite{Lin} and
reconstituted ODAs systems shown in previous studies \cite{Oda}.

The EM images and subsequent analysis also provide information about the
occupancy of dynein arrays on microtubules (see SI text and Figs.~S3 and
S4 for more details).  At high mixing ratios ODA arrays occupy most of
microtubules and the length of the arrays are probably underestimated
due to the limited size of the observation field. However, the length
distribution of contiguous dynein arrays shows that the average length
of about $\sim 200\,\rm nm$ (corresponding to approx.\ eight ODAs) is
largely independent of the mixing ratio (Fig.~\ref{fig:Fig1}C).  We
concluded that the ODA alignment is a highly cooperative assembly
process.

\begin{figure*}
  \centering
  \makebox[\textwidth][c]{\includegraphics[width=1.4\textwidth]{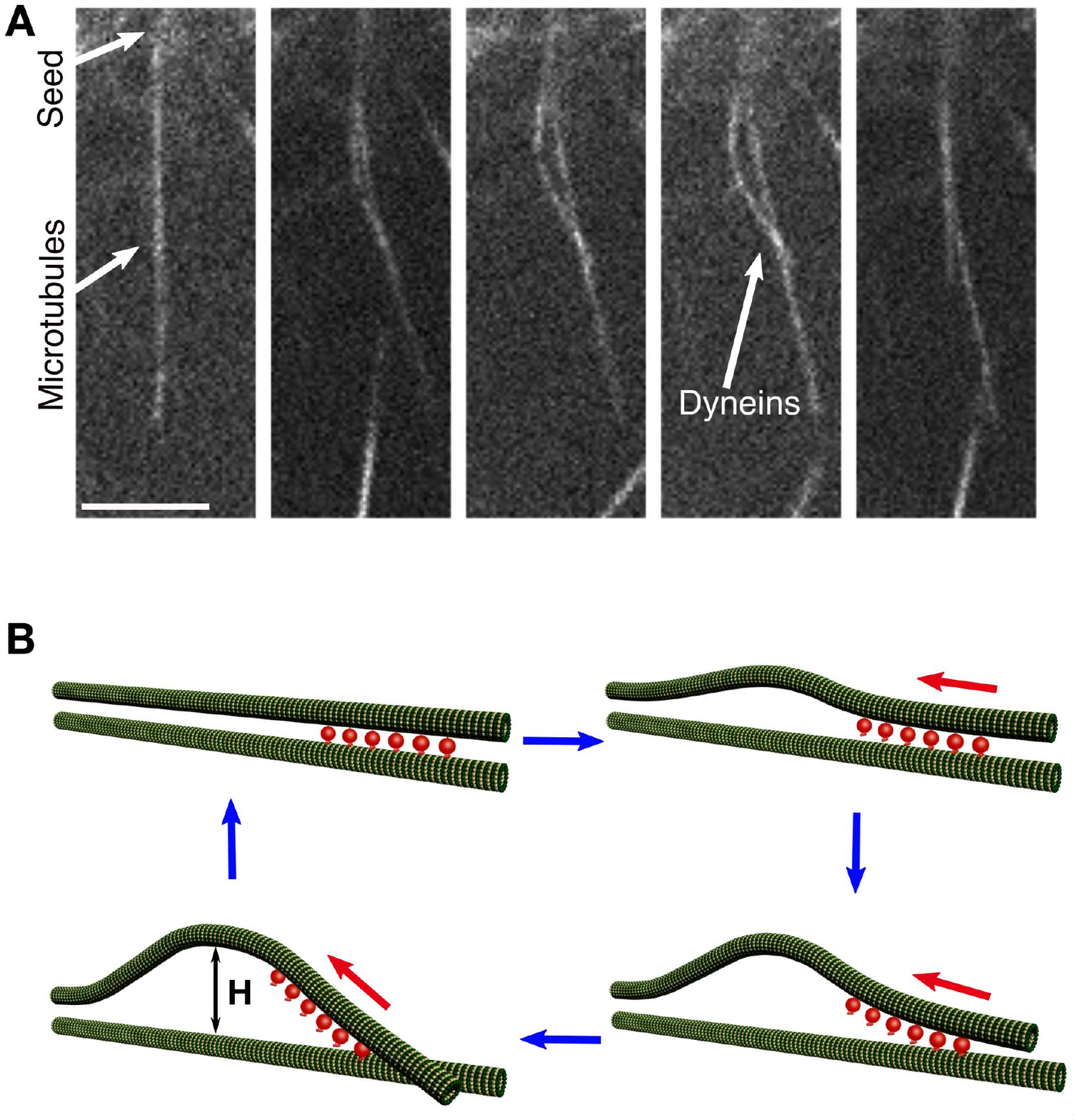}} \caption{A)
    Micrographs of bending cycle of two microtubules forming a
    synthoneme. Scale bar: 10 $\mu\rm m$ B) Schematic representation of
    the filament buckling. Initially, the filaments (green) are straight
    and the dynein motors (red) enter the active state. The force
    produced by the dyneins buckles the filament on which they are
    attached. The buckling changes the crossing angle between the
    microtubules, which reduces the number of dyneins that can bind to
    both of them.}
  \label{fig:Fig2}
\end{figure*}

After adding 1mM Mg--ATP to the system we observe many microtubule pairs
with a joint basal end bending
persistently 
We explain the functional mechanism behind the observation in the
following way. Free microtubule ends fluctuate in the bulk. Dynein
motors assembled on one microtubule attach to the adjacent microtubule,
then Brownian fluctuations are suppressed. By using energy from ATP
hydrolysis, the motors cooperatively produce shearing forces between
these two microtubules (see Fig.~\ref{fig:Fig2}A). When they reach the
buckling instability, the filaments separate.

In more detail, one microtubule slides towards the seed. When the
crossing angle between the microtubules increases, the forces on the
motors increase and thus an increasing number of motors detaches without
the possibility to rebind. The other microtubule bends only weakly. The
motion is stalled when the force produced by the remaining motors
reaches the force of the buckled microtubules. After that, stochastic
fluctuations can tip the motors into cooperative detachment, akin to a
catastrophic failure, which leads to a fast straightening of the
microtubules and a return to the original state. The straight
microtubules again allow the attachment of a larger number of motors. We
observed that this process could repeat itself periodically. This cycle
and the functional mechanism are illustrated in the schematic
representation in Fig.~\ref{fig:Fig2}B.

The long lag phase and subsequent rapid sliding leading to rapid
buckling indicate that the attachment and movement of motors take place
in a cooperative manner, too.  The simplest possible explanation is that
the attachment rate of the first motor between unconnected microtubules
is low because of thermal fluctuations. The first motor capturing the
adjacent microtubule then suppresses the fluctuations and largely
accelerates the attachment of subsequent motors. The mechanism in which
microtubule buckling accelerates the detachment of motors is also at the
core of the geometric clutch hypothesis for ciliary
beating \cite{Lindemann}. On the other hand, the release of the motors
under load is reminiscent of models that propose a dynamic instability
in the sliding motion between microtubule
doublets \cite{Camalet.Prost1999}.

\begin{figure*}
  \centering \includegraphics[width=1\textwidth]{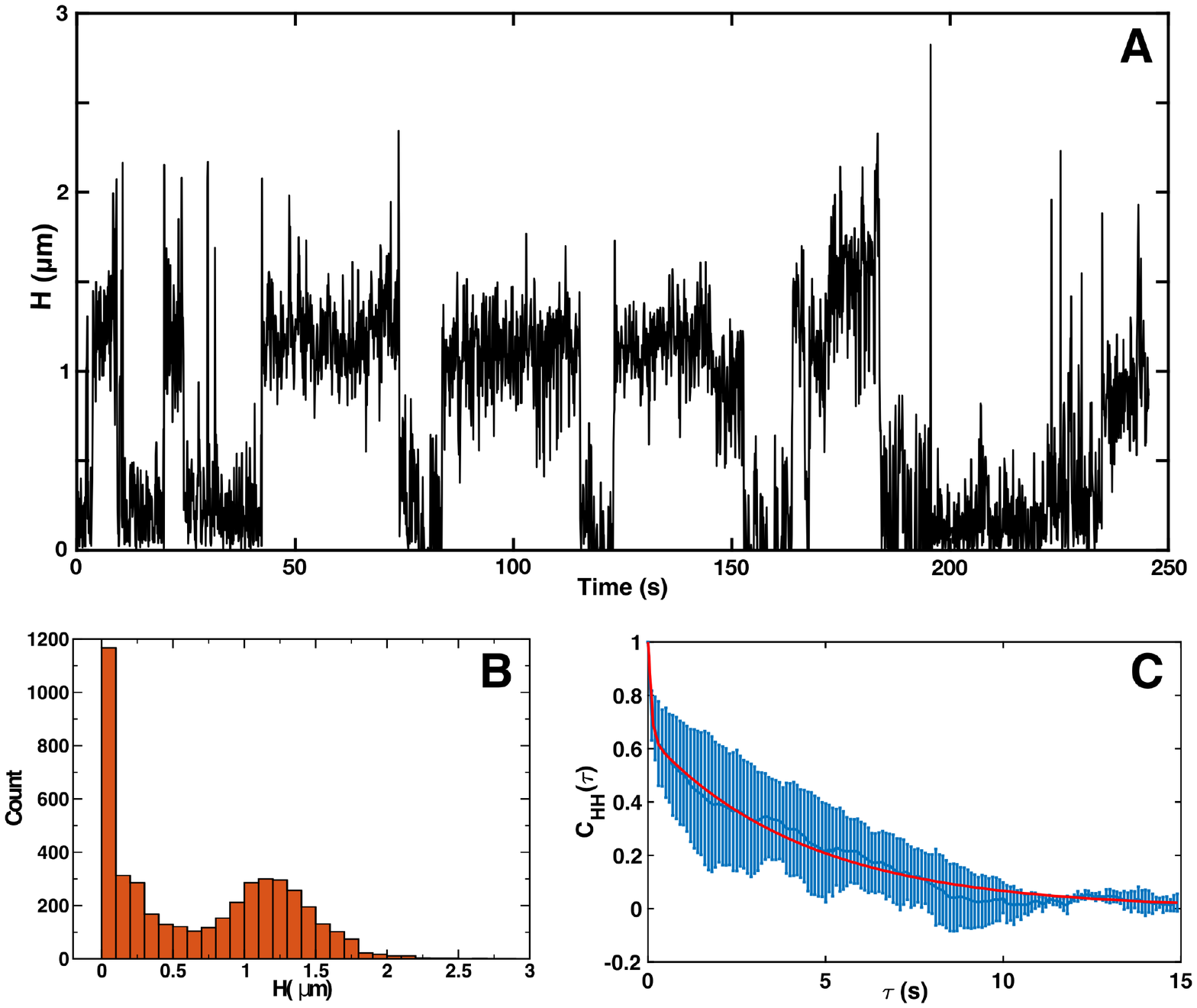} \caption{A)
    Maximal distance between the filaments in the buckled region ($H$),
    as defined in Fig.~\ref{fig:Fig2}, as function of time. B)
    Distribution of distances $H$ between the filaments. As the
    distribution is clearly bimodal, we designated the two states as
    open and closed. C) Normalized autocovariance function
    $C_{HH}(\tau)=(\left< H(t+\tau) H(t)\right>- \left<H\right>^2) /
    \left< (H(t)-\left<H\right>)^2 \right>$.  }
  \label{fig:Fig3}
\end{figure*}

We quantify the oscillations by measuring the maximal distance $H$
between the buckled microtubules. A typical time series representing the
bending cycle is shown in Fig.~\ref{fig:Fig3}A. This cycle can be
divided into four phases consisting of an active bent state and a
relaxed state with variable duration and the two transitions between
them, namely a rapid bending from the relaxed to the active state and a
rapid unbending from the active to the relaxed state. The four phases
are influenced also by the stochastic nature of dynein motors. Both
rapid bending and unbending phases completed within a few hundred
milliseconds. The swiftness of the transitions is a likely clue on the
cooperativity of the motors as the motor recruitment acts collectively
to transition between states.

By analyzing the histogram of distances $H$ in Fig.~\ref{fig:Fig3}B we
can observe a bimodal distribution. Therefore, the above description of
the bending cycle as a process with two states is a suitable one. We
will call them \textit{open} and \textit{close}, and consider stochastic
transitions between them. We denote the rates of the closed-open and
open-closed transition with $k_+$ and $k_-$, respectively. The
discretized variable $H$ then has the character of a telegraph noise
process. A robust way to determine the two rate constants is from the
occupancy ratio ($P_{\rm open}/P_{\rm closed}=k_+/k_-$) and from the
autocovariance function of $H$, which for the telegraph process has the
dependence $\sim \exp(-(k_++k_-)\tau)$. The autocovariance functions are
shown in Fig.~\ref{fig:Fig3}C. They have a bi-exponential form --
besides the slow component, related to the open/closed transitions, a
fast one resulting from noise with a short correlation time. From the
slow rate and $P_{\rm open}/P_{\rm closed}=1.7 \pm 0.5$, we obtain the
rates $k_+=(0.18 \pm 0.07)\,\text{s}^{-1}$ and
$k_-=(0.10 \pm 0.04)\,\text{s}^{-1}$.

We use the elastic model that is described in the next section to
estimate the force and the energy needed to buckle the filaments.  With
a typical length of $l=8\,\rm \mu m$ and $EI=4\times 10^{-24} \rm Nm^2$,
Eq.~(\ref{eq:buckling}) gives $F\approx 1\,\rm pN$, showing that a small
number of dyneins is sufficient to buckle the filaments. For a typical
distance $H=1.2\,\rm \mu m$, we further calculate the elastic energy of
the buckled state as
$F\Delta s\approx 480\times 10^{-21} {\rm Nm}\approx 116\, k_B
T$. Changes in elastic energy that exceed the thermal energy by two
orders of magnitude prove that the beating is actively driven by
molecular motors, i.e., that we assembled an active system.

\subsection{Theory and Simulation.}

\begin{figure*}
  \centering \includegraphics[width=0.7\textwidth]{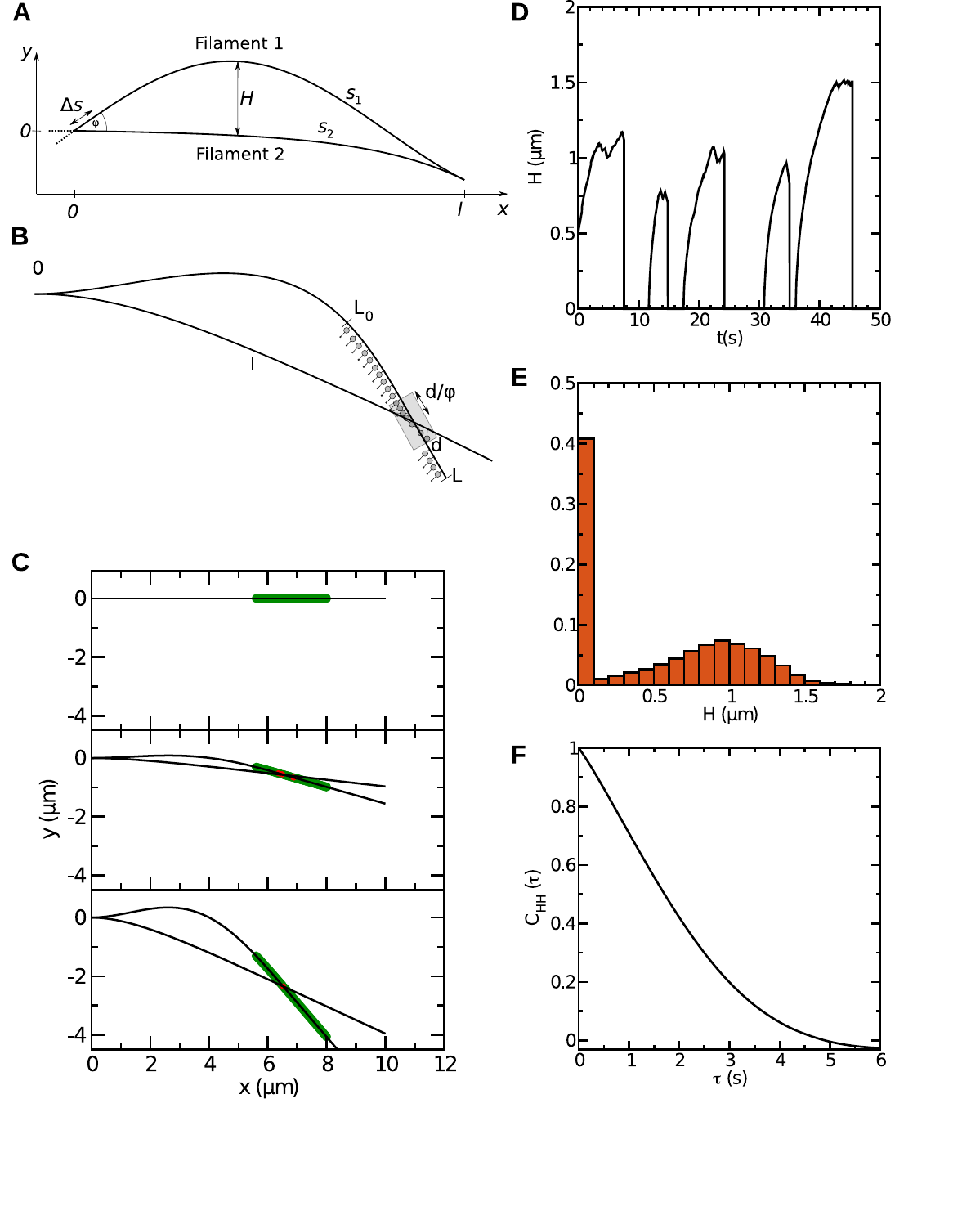}
  \caption{Theoretical description of the active synthoneme. A) Elastic
    model. Two filament segments of different lengths are pinned to each
    other at $x=0$ and clamped at $x=l$. They are parameterized with the
    functions $y_1(x)$ and $y_2(x)$. The $x$-axis represents the
    direction of the force. The condition for the pinned end at $x=0$ is
    $y_1(0)=y_2(0)=0$. At the clamped end, the condition is
    $y_1(l)=y_2(l)$ and $y_1'(l)=y_2'(l)$. B) Active model: two
    filaments are clamped in the origin. Filament 1 is decorated with
    motors from length $L_0$ to length $L$. Motors can bind to filament
    2 if the distance between the filaments is smaller than $d$ (grey
    rectangle). C) Snapshots from a stochastic simulation. The black
    lines show two filaments, red and green circles show bound and free
    motors, respectively. The upper panels shows the filaments before
    binding, the middle panel the onset of buckling and the lower panel
    a stalled state before catastrophic detachment of
    motors. 
    D) The distance between the filaments $H$ as a function of time from
    the simulation. E) Distribution of distances $H$ from the
    simulation. As in the experiment, a strongly bimodal distribution is
    visible. F) Autocorrelation function of the distance $H$ as a
    function of time.}
  \label{fig:Fig4}
\end{figure*}

The shapes of the two bent filaments (Fig.~\ref{fig:Fig4}A) and the
forces that the dyneins need to produce to enter the bent configuration
can be calculated by linear elasticity theory if we assume that the
shape is quasi-static and that the deflections are small.  In contrast
to earlier models \cite{Brokaw2009}, we allow both filaments to
bend. Because both ends are force-free, the force in filament 1 is
exactly opposite to that in filament 2. The curvature in each filament
is proportional to the local bending moment
\begin{equation}
  EI\,y_i''(x)=F_i y_i(x)\;.
\end{equation}
The resulting equation is solved with the ansatz $y_1=A \sin kx$,
$y_2=B \sinh kx$ with $k=\sqrt{F/EI}$. The boundary condition at $l$
leads to the equation $\tan kl=\tanh kl$ with the lowest non-trivial
solution $kl=3.9266$ and
\begin{equation}
  \label{eq:buckling}
  F=1.5622 \frac{\pi^2 EI}{l^2}\;.
\end{equation}
This equation can be used to estimate the force produced by the motors
that is needed to buckle the filaments. Notably, the prefactor in the
buckling load of $1.5622$ lies between the values of $1$ that applies to
the classical Euler problem with two pinned ends and $2.046$ for one
pinned and one clamped end \cite{Gere}. As expected, clamping one end to
a flexible filament leads to a lower critical load than clamping it to a
fixed direction in space.

The length difference between the filament segments, $\Delta s$, and the
distance $H$ can be derived as functions of the crossing angle $\varphi$
(see SI).

\paragraph{\textbf{Dynamical model.}} The behavior of the system over
time observed during the experiments can be described by a dynamical
model. The two filaments (each of length $L$) are clamped at the minus
end (Fig.~\ref{fig:Fig4}B). Filament 1 is bearing dynein motors between
length $L_0$ and $L$, at a line density $\rho$. Each motor can be in the
attached or detached state. Attachment can take place in the overlap
zone $l_1\pm d/\varphi$, where $d$ is a characteristic distance between
filaments that the motors can bridge. In that zone, the attachment rate
of the motors is $k_{\rm on}$. However, we use a lower constant
$k^0_{\rm on}$ for two reasons: (i) a reduced attachment rate for the
motors that align end-to end along one filament indicates the
cooperativity between the motors; (ii) the effective attachment rate is
strongly reduced by thermal fluctuation when the motors are
simultaneously in the detached state. For a motor loaded with force $f$
we use a detachment rate $k_{\rm off}=k_{\rm off}^0 \exp(\alpha f)$,
where $\alpha$ is a parameter representing the inverse characteristic
detachment force.

A motor attached on filament 1 with load $f$ moves towards the minus end
of filament 2 with velocity $v=\dd(\Delta s)/\dd t$ determined by a
linear force-velocity relationship
\begin{equation}
  \label{eq:fv}
  f=f_s(1-v/v_m)
\end{equation}
with the parameters $f_s$ (stall force) and $v_m$ (unloaded velocity).
When $n$ motors are attached, they share the load equally such that
$ F(l_2)=n f $.

We explain the oscillatory cycle of the filaments observed in the
experiments through the following stages: First, all motors are detached
and the first attachment takes place randomly with the rate
$k^0_{\rm on}$. This increases the binding rate of remaining motors to
$k_{\rm on}$. After a short time, the total motor force $n f_s$
surpasses the buckling force $F(l)$, where $l$ is the distance to the
first bound motor. The straight state becomes unstable.  By considering
the duty ratio of stalled motors (fraction of time spent in the attached
state)
\begin{equation}
  \label{eq:duty}
  \eta=\frac{k_{\rm on}}{k_{\rm on}+k_{\rm off}^0 \exp(\alpha f_s)}\;,
\end{equation}
the condition of buckling is
\begin{equation}
  \eta f_s \rho (L-L_0) > F(L_0)= 1.5622 \frac{\pi^2 EI}{L_0^2}\;.
\end{equation}
The initial buckling instability is followed by the growth of the bulge,
driven by the sliding action of motors in the overlap zone. During this
phase, the crossing angle $\varphi$ increases, which reduces the length
and the number of motors in the overlap zone, leading to a metastable
state, where the force balance is reached at an angle $\varphi$ when
$\eta \rho f_s 2d/\varphi = F(l)$.

Through random detachment of motors, a ``catastrophic failure'' can take
place when detachment of motors leads to an increased load on the
remaining ones, further increasing their detachment rate. We assess the
stability of a state with the following consideration.  At a constant
total force $F$, initially distributed over $n$ attached motors, the
on-off dynamics is given by the master equation
\begin{equation}
  \dot n=(N-n) k_{\rm on}-n k_{\rm off}e^{\alpha F/n}. 
\end{equation}
Its stability is determined by the derivative
\begin{equation}
  \frac{\partial\dot n}{\partial n}=- k_{\rm on} +\left(\frac{\alpha F}{n}-1\right) k_{\rm off}e^{\alpha F/n} 
\end{equation}
which is positive when $\alpha f_s n_0 / n > 1/( 1-\eta)$.  As a rough
estimate, we expect an instability when $\alpha f_s \gtrsim 1$.

Through full detachment, the filaments finish in the straight position
where new motors quickly re-attach and repeat the cycle. This is
compatible to the cyclic dynamics that we hypothesized for our
experimental outcomes.

The elastic model and the motor model described above are combined in a
stochastic simulation that confirms the experimental results
(Fig.~\ref{fig:Fig4}C-F). The distance between the filaments oscillates
with steep transitions between the open and the closed state
(Fig.~\ref{fig:Fig4}D), closely resembling the experimental traces shown
in Fig.~\ref{fig:Fig3}A. The histogram of distances shows a clear
bimodal distribution (Fig.~\ref{fig:Fig4}E), as in the experimental data
Fig.~\ref{fig:Fig3}B. Finally, the autocorrelation function shows a
near-exponential decay (Fig.~\ref{fig:Fig4}F). In contrast with the
experiment (Fig.~\ref{fig:Fig3}C), the fast phase is missing as we did
not include thermal noise and measurement uncertainty in the simulation.

\section*{Conclusion}

The reconstituted minimal axoneme, i.e. the synthoneme, that we present
in this study is a one-of-a-kind result resembling the natural system
with its beating behavior.  All crucial components of our synthoneme are
assembled in-vitro in a bottom-up approach. We show that the
cooperativity of the dynein motors to arranging on the filament as well
as to actuating the buckling can be reconstituted in-vitro.  Although
the amplitude and frequency of oscillations are still well below those
of natural cilia, we have shown that the basic functionality can be
established with a far smaller number of proteins. Specifically, while
cilia contain several hundred different proteins, our synthetic axoneme
works with only three components (tubulin, ODA, ODA-DC).  The
observation that synthetic systems consisting only of molecular motors
and filaments can reproduce the basic movement necessary for
cilia/flagella beating also provides clues about the amount of
complexity needed by the ancestor cilia to first achieve beating
motility. Indeed, some organisms like the male gamete of the parasitic
protozoan \textit{Diplauxis hatti} have a motile axoneme with only three
doublets instead the classical ``9 + 2'' structure \cite{Prensier}. This
suggests that bending behavior could be produced by simpler structures
that have filaments and motors.  Future refinements will lead to a
convergence between the synthetic and the natural cilium. This will
provide a systematic way of dissecting the elusive beating mechanism.

Besides the potential of this system for answering important questions
about ciliary beating, our synthetic cilium may encourage the
technological development of molecular machines for fluid transport at
micro- and nanoscale.  Most previous attempts to build artificial cilia
have concentrated on magnetic \cite{Vilfan.Babic2010,Dong.Sitti2020},
electrostatic \cite{Toonder.Anderson2008} or
optical \cite{vanOosten.Broer2009} actuation mechanisms to produce motion
similar to the natural one. These studies showed hydrodynamic
entrainment between adjacent structures and creation of metachronal
waves. However, these systems are unrelated to biological cilia. In the
future our system will also be developed as multi-cilia system by using
axoneme seeds as the base and aligning them and we will achieve the
synchronization of axonemal beating.

\section{Materials and Methods}
\subsection{Preparation of the crude dynein extract}
Demembranated axonemes are suspended in 0.6M KCl containing HMDEK
solution to extract crude dynein sample (see SI for more details about
preparation). After spinning down the axonemes for 5 min, the
supernatant is desalted with the overnight dialysis against HMDEK
solution (Spectra/Por Dialysis Membrane, MWCO:100-500D, Biotech CE). The
concentration of the crude dynein extract is measured with Bradford
method and the concentration used during the experiments was 0.02 mg/ml.
	
\subsection{Preparation of dynein–MT complexes}
The flow chamber was build with Teflon-treated coverslips as previously
described in \cite{Torisawa} and spaced with double-coated tape (80mm
thick, W-12; 3M, St.~Paul, MN) to prevent any nonspecific binding of
proteins onto the surface. Microtubules growing close to each other and
with the same polarity were obtained by using fragments of demembranated
axonemes prepared by rigorous pipetting as polymerization seeds.  The
ODA–MT complexes are reconstituted in the flow cell by flowing the
components one after the other.  Briefly, in the flow cell a small
amount of axonemes are attached on the bottom of the flow cell.  After
5-min incubation, the flow cell is washed with 1\% (w/v)
Pluronic\textsuperscript{\textregistered} F127 in BRB80 (80 mM PIPES,
1~mM MgCl$_2$, 1 mM EGTA, pH 6.8 with KOH) and is incubated for 5
min. After washing the chamber with BRB80, fluorescently-labeled
(Cy3--labeled) porcine tubulin (3\% labeling) is introduced into the
flow cell, polymerized in the presence of 1 mM GTP, 50\% DMSO, 1mM
MgCl$_2$ at 37$^{\circ}$C for 30 min and stabilized with 7 $\mu$M
taxol. After microtubule polymerization, diluted crude-dynein extract is
introduced into the flow cell and incubated for 5 min. The non-bound
protein is eliminated by washing the chamber with buffer and afterwards
1mM ATP is perfused into the chamber to trigger the activity.
	
\subsection{Imaging and tracking}
	
Fluorescence images of the MT-ODA complex are acquired using an inverted
fluorescence microscope Ti-E (Nikon, Japan) equipped with a 60$\times$
CFI Apochromat objective (N.A.=1.49, Nikon, Japan) and the confocal unit
(CSU-X1, YOKOGAWA, Japan). The data is recorded with an iXon Ultra EMCCD
camera (Andor Japan, Japan). The images are acquired at a frequency of
10 Hz.  The movement of the filaments over times was tracked manually
and a third order polynomial was fitted to the data by using a
purpose-written MATLAB code.

\paragraph{\textbf{Acknowledgements}}
E.B., I.G., and R.G. acknowledge support from the MaxSynBio
Consortium which is jointly funded by the Federal Ministry of
Education and Research of Germany and the Max Planck
Society. E.B. acknowledges support from the Volkswagen Stiftung
(priority call ``Life?''). A.V. acknowledges support from the
Slovenian Research Agency (grant no.\
P1-0099). K.O. acknowledges MEXT/JSPS KAKENHI, grant numbers
26440089, 17K07376 and JP16H06280, the Takeda Science Foundation
and Hyogo Science and Technology Association that partly
supported the project.
        
\bibliographystyle{naturemagNoURL}
\bibliography{Reference_TF}

\setcounter{equation}{0} \setcounter{figure}{0}
\renewcommand{\theequation}{S\arabic{equation}}
\renewcommand{\thefigure}{S\arabic{figure}}

\clearpage
\section*{Supplementary Materials and Methods}

\subsection{Preparation of \textit{Chlamydomonas reinhardtii}. Axonemes and their demembranation.}
Axonemes were obtained from \textit{Chlamydomonas reinhardtii} according
to the dibucaine method \cite{Kagami,Witman1978}. \textit{C. reinhardtii}
wild type strain (137c mt-) and the outer-arm less mutant
(\textit{oda1}) were cultured in 4 liters of the liquid TAP medium under
continuous illumination at 20 $^{\circ}$C for 4 days with air
bubbling. Cells were harvested by centrifugation at 1692$\times$g (3000
rpm, R10A3 rotor, Himac CR22E) for 6 min, and then re-suspended in 40 ml
ice-cold HMDS (30 mM HEPES-NaOH, 5 mM MgSO$_4$, 1 mM dithiothreitol
(DTT), 4\% sucrose, pH 7.4). All procedures described below were
conducted at 4 $^{\circ}$C or on ice unless otherwise stated. The cell
suspension was moved to a 50-ml conical centrifuge tube and 2.5 mM
(final concentration) dibucaine HCl (Wako, 167-15111) was added. The
suspension was pipetted rapidly in and out of 10-ml Falcon disposable
plastic pipettes for 30 s. Afterwards 200 $\mu$l of 100 mM EGTA was
added and the solution was gently resuspended several times. The cell
bodies were removed by centrifugation at 1670$\times$g (3000 rpm, RS-4
rotor, KUBOTA 6800) for 6 min, the supernatant was transferred to a new
50-ml conical centrifuge tube and centrifuged again as above.  The
cell-free supernatant was centrifuged at 27720$\times$g (15,000 rpm,
RSR20-2 rotor Himac CR21) for 12 min. The resulting pellet was
resuspended in 1.5 ml of HMDEK (30 mM HEPES-NaOH, 5 mM MgSO$_4$, 1 mM
DTT, 1 mM EGTA, 50 mM potassium-acetate, pH 7.4) and demembranated by
adding 15 $\mu$l of 20\% Nonidet P-40 (NP-40, Nacalai Tesque). After 5
min-incubation, the suspension was centrifuged at 21130$\times$g (15,000
rpm, FA-45-24-11 rotor, Eppendorf 5452R) for 5 min, and resuspended in
1.5 ml HMDEK. This procedure was repeated twice to remove
detergent. Axonemes were collected by centrifugation as described above.
An aliquot of the demembranated \textit{oda1} axonemes was used to
prepare the seeds of microtubule polymerization. Suspension of the
axonemes was gently homogenized with a Teflon pestle and pipetting. The
resultant suspension was centrifuged at 3380$\times$g at 4 $^{\circ}$C
for 10 min. The pellet was resuspended with a small amount of HMEDK
solution. Intrinsic dyneins of the seeds were deactivated and the
sliding among doublet microtubules was inhibited with the treatment of
ethylene glycol bis(succinimidyl succinate) (Thermo Fisher Scientific,
21565).
	
\subsection{Preparation of of the crude dynein extract.}
For the extraction of ODAs and docking complex from the
\textit{Chlamydomonas reinhardtii} wild-type axonemes we follow the
method described in previous works \cite{Haimo,kamiya2005}.  As
reference, we prepared high-salt extract from \textit{oda1} (outer-arm
less mutant) axonemes.  Demembranated axonemes were resuspended in 0.6M
KCl containing HMDEK solution to extract crude dynein sample. After
spinning down the axonemes at 21130$\times$g (15,000 rpm, FA-45-24-11
rotor, Eppendorf 5452R) for 5 min, the supernatant was desalted with the
overnight dialysis against HMDEK solution (Spectra/Por Dialysis
Membrane, MWCO:100-500D, Biotech CE). The concentration of the crude
dynein extract was measured with Bradford method and it was around 0.02
mg/ml.  High-salt extract from the demembranated axonemes contains
various types of proteins \cite{Haimo}. Therefore, protein components of
the dynein-microtubule complex should be examined with SDS-PAGEs.
Spinning down the crude dynein extract with microtubules in the presence
of 1mM ATP partially pulled down the outer-dynein arms and related
proteins with microtubules. Using \textit{oda1} axonemes, we repeated
the same procedure on the axonemes and high-salt extracts. The resultant
supernatants and pellets of the centrifugations were examined with the
SDS-PAGE and the contents of \textit{oda1} were compared with those of
wild type.  Inner arm dyneins, components of the central pair apparatus
and radial-spoke components remain in supernatant under this condition
(Fig.\ S1). Thus, our procedure provided ODA-DC-MT complex which
contained no inner arm dyneins but only outer arm dyneins.
	
\subsection{Negative staining electron microscopy on dynein-MT complex.}
To evaluate the formation of ODA arrays on a microtubule, we examined
the structure of dynein-MT complex with an electron
microscope. Structural similarity of this complex to the axoneme has
been reported in previous studies and was examined by negative staining
electron microscopy. It showed that microtubules were periodically
decorated with ODA dyneins. The crude dynein extract of 200 $\mu$g/ml
was mixed with taxol-stabilized microtubules (200 $\mu$g/ml) at the
various ratios (10:1, 1:1, 1:2 and 1:10). After 30-min incubation, the
small aliquot of the mixture (5 $\mu$l) was applied on to a carbon grid
and washed with MMEK for EM (30 mM MOPS-NaOH, 5 mM MgSO$_{4}$, 1 mM DTT,
1 mM EGTA, 50 mM KCl, pH 7.4). Fixation with 2\% glutaraldehyde, washing
with MMEK and staining with 1.4\% uranyl-acetate with 100 $\mu$g/ml
bacitracin. The sample was examined with JEM-1400 (JEOL, Japan) with
80kV acceleration voltage and the images of the samples were collected
with 4K camera (US4000, Gatan). Typical electron micrographs are shown
in Fig.~S3.  Coverage of the microtubules with ODA-DC complex was
evaluated with image analysis. To extract the 24 nm structural repeat of
dynein-ODA-DC from the electron micrographs, the custom-built band-pass
filter was applied to them. The merger of the original and filtered
image clearly shows the presence of dynein-DC complex. The length of the
complex and microtubule length in the image field were measured.

\section*{Elastic model}
	
In this section we derive the relationships between the crossing angle
$\varphi$, the length difference $\Delta s$ and the distance between
filaments $H$ (Fig.~3A in main text). We start with the solution derived
in the main text, where the filaments shapes are described by
$y_1=A \sin kx$ and $y_2=B \sinh kx$, with $kl=3.9266$. The amplitude
ratio is determined from the condition
\begin{equation}
  \frac A B = \frac{\sinh kl}{\sin kl}= -0.02785\;.
\end{equation}
In the linear approximation, the opening angle at the pinned end is
\begin{align}
  \varphi&=y_1'(0)-y_2'(0)=Ak-Bk=Ak\left(1-\frac{\sin kl}{\sinh kl}\right)\notag \\ &=1.02785\, Ak\;.
\end{align}
Also in the linear approximation, the contour length of filament $i$
($i=1,2$) can be calculated as
\begin{equation}
  s_{i}=\int_0^l \sqrt{1+(y_{i}')^2} \dd x\approx \int_0^l\left[1+\tfrac12 (y'_{i})^2\right] \dd x
\end{equation}
and the difference as
\begin{align}
  \Delta s&=\int_0^l\left[\tfrac12 (y'_1(x))^2 - \tfrac12 (y'_2(x))^2 \right]\dd x=\frac {lk^2}{4}(A^2-B^2)\notag\\ &=\frac {\sinh kl +\sin kl}{\sinh kl -\sin kl}\frac {l \varphi^2}{4}=0.2364\, l \varphi^2\;.
\end{align}
Finally, we define the maximum distance between the filaments
$H=\max(y_1-y_2)$. The condition for maximality is
$y_1'(x_m)-y_2'(x_m)=0$ and leads to
\begin{equation}
  \frac{\cos kx_m}{\cosh kx_m}= \frac{\cos kl}{\cosh kl}
\end{equation}
with the solution $k x_m=1.6458$ and
$H=y_1(x_m)-y_2(x_m)=0.24301\, l \varphi$. These relationships are used
in the main text to estimate the elastic energy of the two-filament
system. They are also needed for the stochastic simulation.

\begin{figure*}
  \centering \includegraphics[width=1\textwidth]{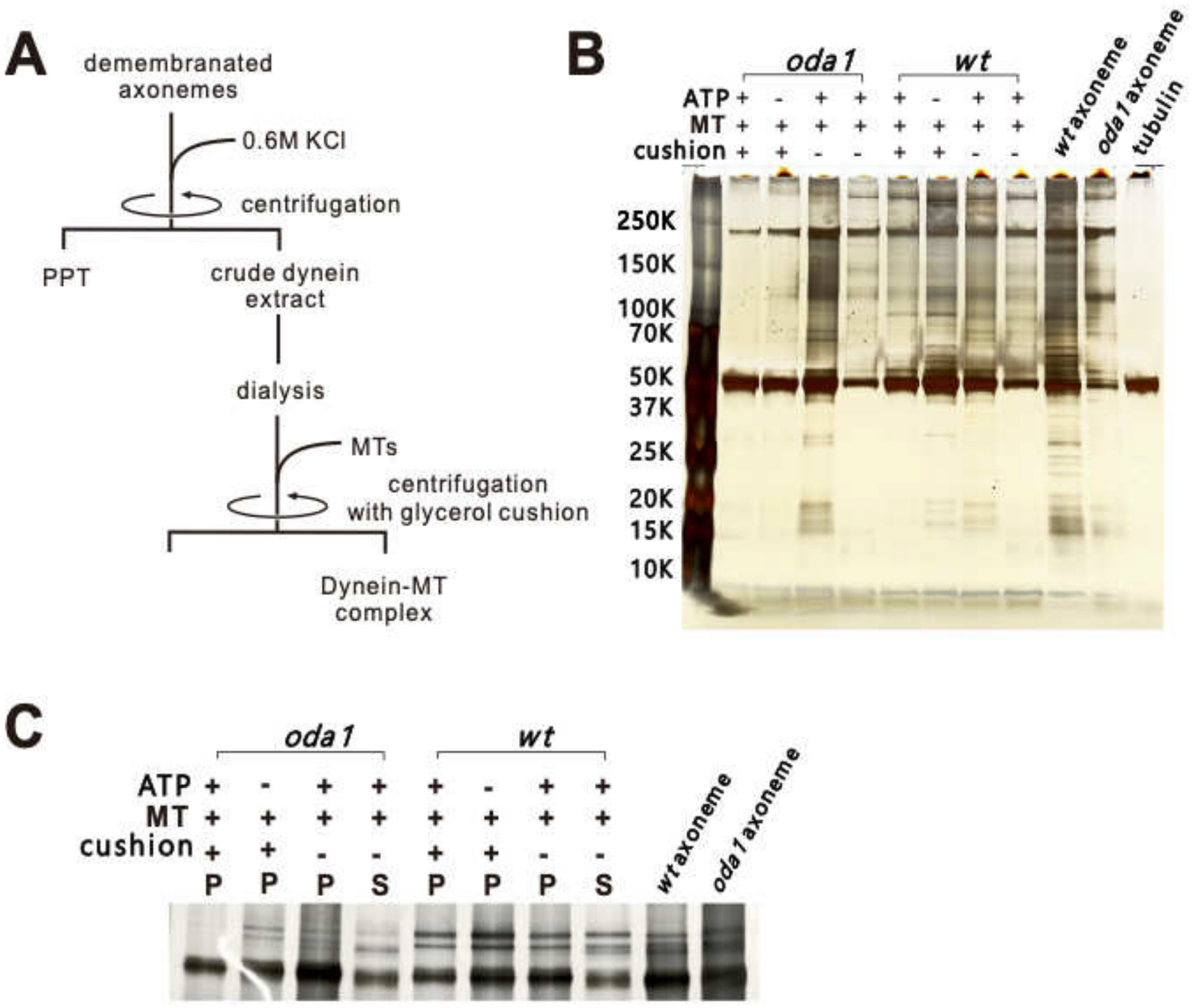}
  \caption{Preparation of the outer-dynein arm-microtubule complex. A)
    Procedure of dynein-crude extraction. PPT means a precipitate B) The
    high-salt crude dynein extract from \textit{Chlamydomonas
      reinhardtii} wild type axonemes, \textit{oda1} axonemes and
    centrifuged dynein-microtubule complex were analyzed by SDS-PAGE
    with silver staining. Since only inner arm dyneins exist in
    \textit{oda1} axonemes, the high salt extract did not contain outer
    arm dyneins. In contrast, the extract from wild-type axonemes
    contains both outer-arm dynein and inner-arm dyneins. The glycerol
    cushion in the centrifugation can separate inner-arm dynein from
    microtubules in the presence of ATP but outer arm dyneins stayed
    with microtubules. Lane 1: molecular weight markers. Lane 2: no
    dynein exists in \textit{oda1}-extract-MT complex. No light chains
    and Intermediate chains of inner-arm dyneins were found. Lane 3:
    Removal of ATP from the complex pulled inner-arm dyneins of
    \textit{oda1} extract into pellet with microtubules. Lane 4: Inner
    arm dyneins were found in the pellet when being centrifuged without
    cushion. Lane 6: Dyneins found in the pellet even in the presence of
    ATP and cushion but no light chains and Intermediate chains of
    inner-arm dyneins were found. This observation suggests that only
    outer arm dyneins remained on microtubules and these dyneins formed
    the complex with microtubules via dynein-docking complex. C)
    High-molecular region of 3\% SDS-PAGE. Lane 1: The microtubule
    pellet in the presence of ATP and glycerol cushion. No dynein heavy
    chain was found in the pellet. Lane 2: Inner arm dynein heavy chains
    were found. Lane 5: Even in the presence of ATP, only outer arm
    dyneins were associated with microtubules. These observations
    suggest that our procedure enabled the ODA-DC-MT complex to contain
    no inner arm dyneins but only outer arm dyneins. S and P indicate a
    precipiate and supernatant, respectively.}
  \label{fig:FigS1}
\end{figure*}
	
\begin{figure*}
  \centering \includegraphics[width=0.8\textwidth]{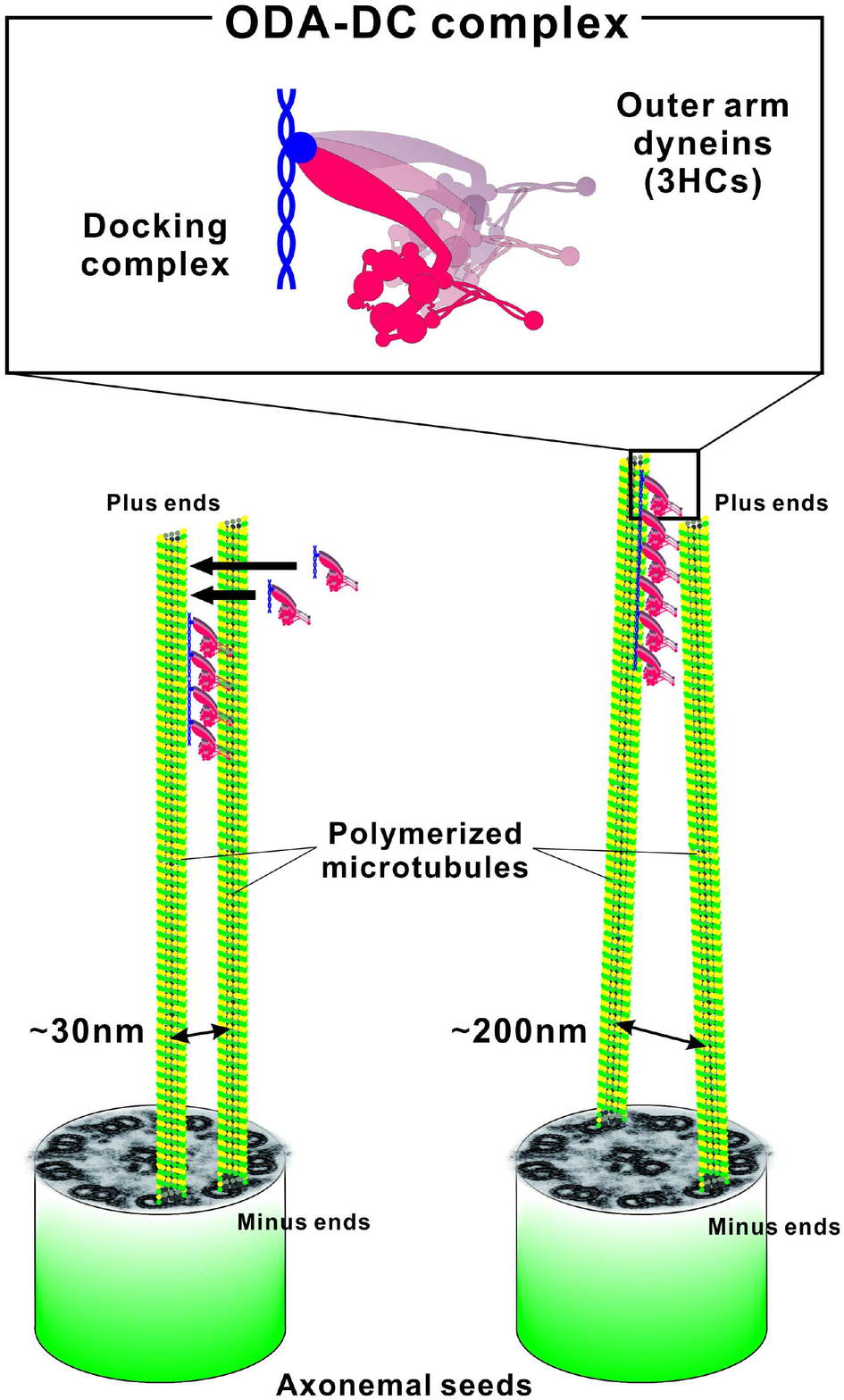}
  \caption{Schematic representation of the assembly of a minimal
    synthetic axoneme with the minimal (a) and maximal (b) distance
    between the polymerized microtubules.}
  \label{fig:FigS2}
\end{figure*}
	
\begin{figure*}
  \centering \includegraphics[width=1\textwidth]{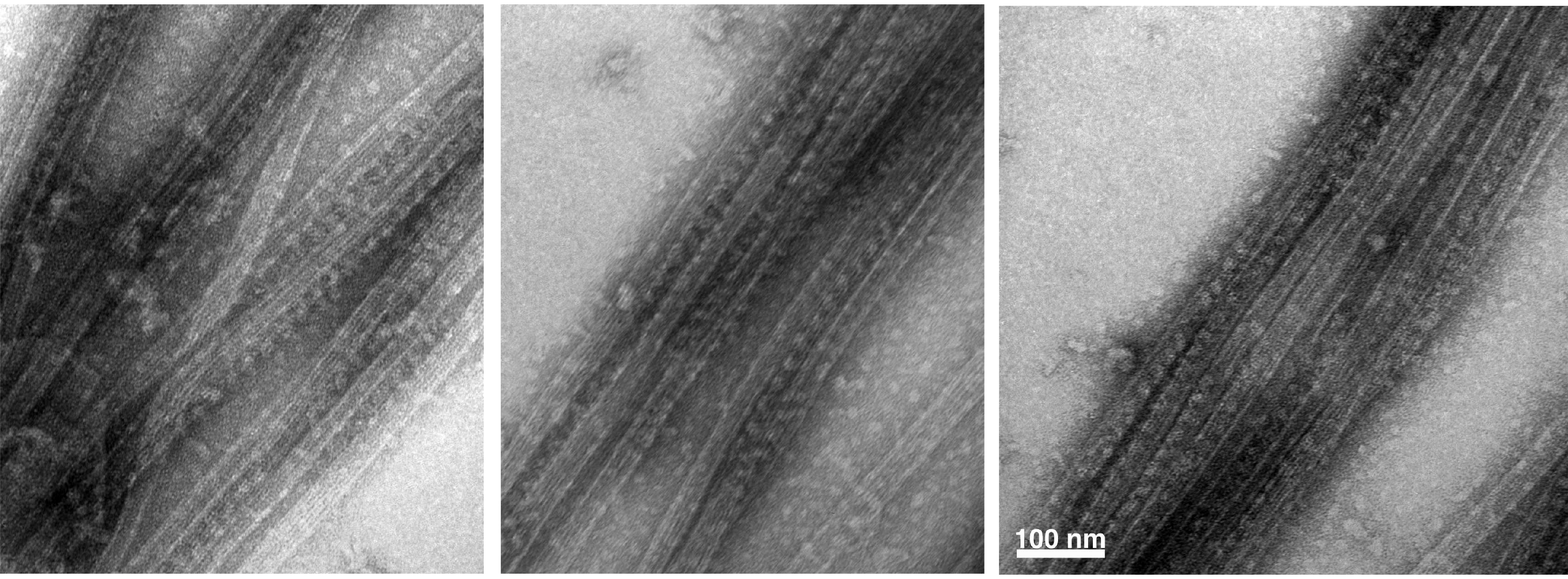}
  \caption{Taxol-stabilized microtubules were mixed with crude dynein
    extract at various mixing ratios and observed with negative staining
    electron microscopy (acceleration voltage 80kV, magnification 25k).}
  \label{fig:FigS3}
\end{figure*}
	
\begin{figure*}
  \centering \includegraphics[width=1\textwidth]{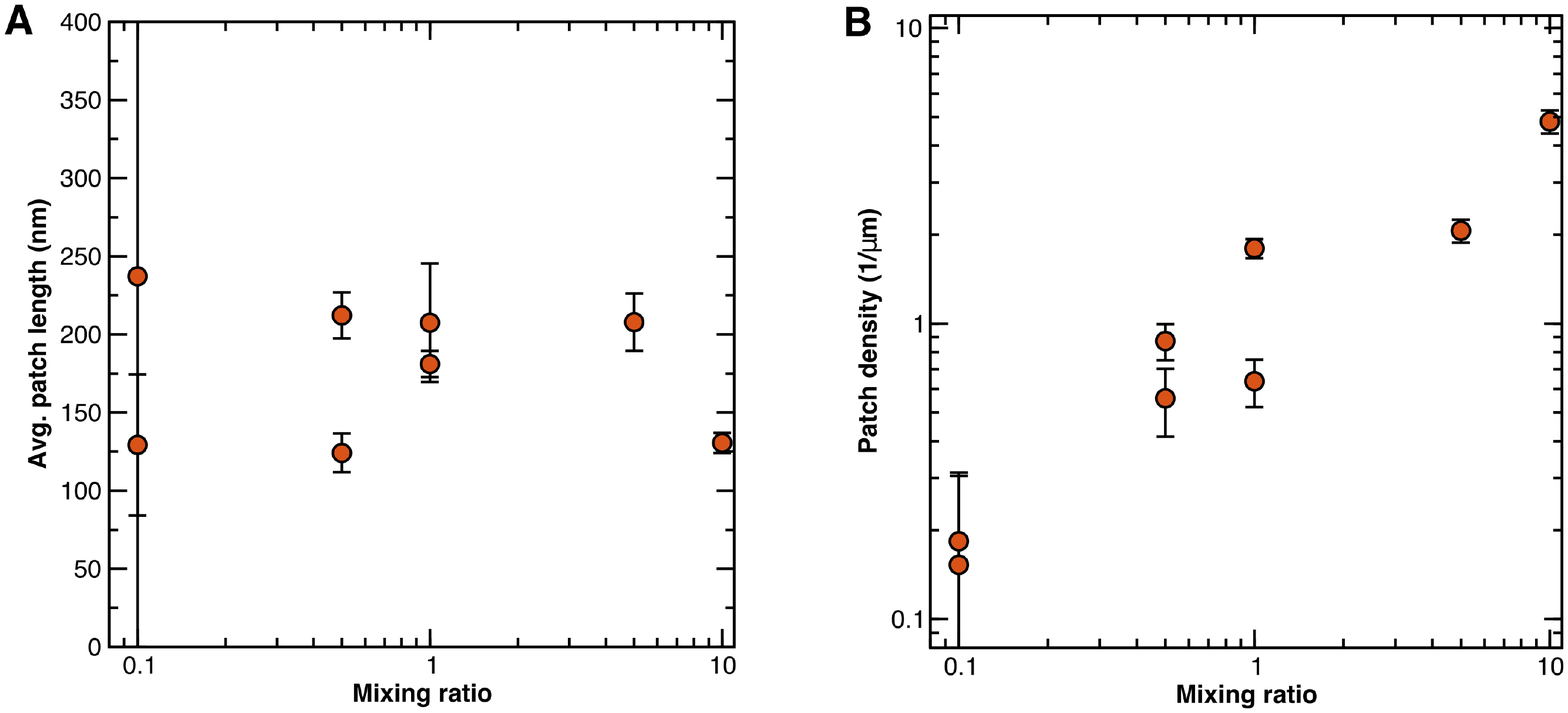}
  \caption{Average length (A) and density (B) of dynein patches on
    microtubules as a function of mixing ration of crude dynein extract
    (4 - 400 $\mu$g/ml) to microtubules (40 $\mu$g/ml).}
  \label{fig:FigS4}
\end{figure*}
\end{document}